\documentclass[showpacs,twocolumn,preprintnumbers,amsmath,amssymb,prl,epsf,superscriptaddress]{revtex4-1}
\usepackage{graphicx}
\usepackage{nicefrac}
\usepackage[squaren]{SIunits}

\begin{document}
\title{Macroscopic Bell States and the Higher-Order Degree of Polarization}
\author{Timur Sh.~Iskhakov}
\affiliation{Max Planck Institute for the Science of Light,
G\"unther-Scharowsky-Stra\ss{}e 1/Bau 24, 91058 Erlangen, Germany}
\author{Ivan~N.~Agafonov}
\affiliation{Department of Physics, M.V.Lomonosov Moscow State University, \\ Leninskie Gory, 119991 Moscow,
Russia}
\author{Maria~V.~Chekhova}
\affiliation{Max Planck Institute for the Science of Light, G\"unther-Scharowsky-Stra\ss{}e 1/Bau 24, 91058
Erlangen, Germany} \affiliation{Department of Physics, M.V.Lomonosov Moscow State University, \\ Leninskie Gory,
119991 Moscow, Russia}
\author{Georgy~O.~Rytikov}
\affiliation{Department of Physics, M.V.Lomonosov Moscow State University, \\ Leninskie Gory, 119991 Moscow,
Russia}
\author{Gerd Leuchs}
\affiliation{Max Planck Institute for the Science of Light, G\"unther-Scharowsky-Stra\ss{}e 1/Bau 24, 91058
Erlangen, Germany} \affiliation{University of Erlangen-N\"urnberg, Staudtstrasse 7/B2, 91058 Erlangen, Germany}
\vspace{-10mm}
\pacs{42.50.Lc, 03.65.Ud, 42.25.Ja, 42.50.Dv}

\begin{abstract}
We study the polarization properties of the macroscopic analogues of two-photon Bell states. The effect of hidden polarization, i.e., polarization of second (and higher) orders in the intensity is observed for the triplet states while the singlet state is demonstrated to be nearly unpolarized at least up to the fourth order. A degree of polarization of arbitrary order is introduced and applied to the macroscopic Bell states.
\end{abstract}
\vspace{5mm}
\maketitle \narrowtext

The difference between polarized and unpolarized light is traditionally described by the degree of polarization (DP)~\cite{Shurcliff}. The commonly used definition for the DP involves only observables that are first-order in the intensity, or second-order in the field,
\begin{equation}
P_1\equiv\frac{\sqrt{\langle S_1\rangle^2+\langle S_2\rangle^2+\langle S_3\rangle^2}}{\langle S_0\rangle},
\label{DP1}
\end{equation}
where $\langle S_j\rangle,\,\,j=0,1,2,3$, are the Stokes parameters~\cite{Shurcliff}. One can show that the first-order DP can be also written in the form
\begin{equation}
P_1\equiv\frac{I_{max}-I_{min}}{I_{max}+I_{min}},
\label{DP1vis}
\end{equation}
where $I_{max},\,I_{min}$ are the maximum and minimum intensities
 observed by transmitting light through a polarizing prism and performing arbitrary polarization transformations before the prism~\cite{Klyshko}. In other words, DP1 coincides with the visibility of the polarization intensity modulation. Definition (\ref{DP1}) does not take into account the second-order moments of the Stokes observables and therefore fails to describe certain effects. It is sometimes referred to as the `classical' DP; however, it is more correct to call it the `first-order' one, as it fails to describe some classical effects as well. In particular, one should mention here 'hidden polarization'~\cite{hidden,Klyshko}: some states can be unpolarized in the first order with respect to intensity but the second-order intensity moments reveal polarization dependence.

Unlike for the first-order DP, which is defined in a unique way by Eqs.~(\ref{DP1},\ref{DP1vis}), various definitions have been proposed for the second-order one. Probably the first attempts were by Chirkin
et al.~\cite{AlArCh} and by Klyshko~\cite{Klyshko}. These definitions are operational ones, allowing
for easy testing in experiment. Since then, several theoretical definitions for DP appeared~\cite{SS,Luis},
and another operational one, introduced by Klimov et al.~\cite{Vik}.
%It described well some nonclassical states studied in experiments.
%Note that for all states considered in Ref.~\cite{Vik}, the first-order DP is close to unity.

Here we consider nonclassical states of light that are not polarized in the first order in the intensity, i.e., their first-order DP is zero. These are `macroscopic Bell states', first theoretically considered in~\cite{Karas} and experimentally produced in~\cite{Macrobell}. Physically, the four states are different. The three triplet ones manifest hidden polarization and the singlet one one does not: it is unpolarized in all orders in the intensity. From this viewpoint, it is interesting to apply to these states the higher-order degrees of polarization.

The Klyshko second-order degree of polarization is defined by analogy with the first-order one, as the visibility of second-order polarization interference~\cite{Klyshko},
\begin{equation}
P_2\equiv\frac{\mu_{max}-\mu_{min}}{\mu_{max}+\mu_{min}},
\label{vis}
\end{equation}
where $\mu$ is some observable given by second-order~\cite{second-order} intensity polarization moments. Depending on the experiment, it can be the variance of a Stokes observable or the second-order Glauber's correlation function. In our experiment, it was the variance of the Stokes observables measured, hence we assume $\mu\equiv\Delta S_{\mathbf{n}}^2$. Here, an arbitrary Stokes observable $S_\mathbf{n}$~\cite{Karas2} corresponds to the operator $\hat{S}_\mathbf{n}$, defined in terms of Stokes operators, $\hat{S}_{1,2,3}$, and a unit vector on the Poincare sphere, $\mathbf{n}\equiv\{\cos\theta;\sin\theta\cos\phi;\sin\theta\sin\phi\}$, with $\phi,\theta$ being the spherical coordinates:
\begin{equation}
\hat{S}_{\mathbf{n}}\equiv\hat{S}_1\cos\theta+\hat{S}_2\sin\theta\cos\phi+\hat{S}_3\sin\theta\sin\phi.
\label{Stokes}
\end{equation}
Note that the quantum observable $\hat{S}_{\mathbf{n}}$ is equivalent to any of the commonly used Stokes operators $\hat{S}_{1,2,3}$; in particular, one can define its mean value, $\langle S_{\mathbf{n}}\rangle$, and the variance, $\Delta S^2_{\mathbf{n}}\equiv\langle(S_{\mathbf{n}}-\langle S_{\mathbf{n}}\rangle)^2\rangle$.

Thus, following the notation of Ref.~\cite{Vik}, we can write the second-order DP definition for the case of our experiment as
\begin{equation}
P_2\equiv\frac{\hbox{supp}_{\mathbf{n}}\Delta S^2_{\mathbf{n}}-\hbox{inf}_{\mathbf{n}}\Delta S^2_{\mathbf{n}}}{\hbox{supp}_{\mathbf{n}}\Delta S^2_{\mathbf{n}}+\hbox{inf}_{\mathbf{n}}\Delta S^2_{\mathbf{n}}},
\label{DP_Klyshko}
\end{equation}
where the maximal (supremum) and minimal (infimum) values of the variance are searched over all possible $\mathbf{n}$ directions.

%The Klimov et al.~\cite{Vik} DP is introduced in a different way,
%\begin{equation}
%\tilde{P_2}\equiv\sqrt{1-\hbox{inf}_{\mathbf{n}}\frac{(\Delta S_{\mathbf{n}})^2}{\frac{1}{3}\langle\mathbf{\hat{S}}^2\rangle}},
%\label{KSS}
%\end{equation}
%where
%\begin{equation}
%\mathbf{\hat{S}}^2\equiv\hat{S}_1^2+\hat{S}_2^2+\hat{S}_3^2.
%\label{SV}
%\end{equation}

The experiment on the generation and measurement of macroscopic Bell states is described in detail in Ref.~\cite{Macrobell}. The states have the form
\begin{eqnarray}
|\Psi^{(\pm)}_{mac}\rangle=e^{\Gamma(a_{1}^{\dagger} b_{2}^{\dagger}\pm
b_{1}^{\dagger}a_{2}^{\dagger})+\hbox{h.c.}}|\hbox{vac}\rangle,\nonumber\\
|\Phi^{(\pm)}_{mac}\rangle=e^{\Gamma(a_{1}^{\dagger} a_{2}^{\dagger}\pm
b_{1}^{\dagger}b_{2}^{\dagger})+\hbox{h.c.}}|\hbox{vac}\rangle, \label{state}
\end{eqnarray}
where $\Gamma$ is the parametric gain coefficient and $a^{\dagger},b^{\dagger}$ are photon creation operators in the horizontal and vertical polarization modes. The subscripts denote frequency (wavelength) modes; in our experiment, $\lambda_1=635$ nm, $\lambda_2=805$ nm. The states can be rewritten in the form of the Fock-state expansion, which allows one to see the photon-number correlations between different polarization modes;
\begin{eqnarray}
|\Psi^{(\pm)}_{mac}\rangle=\sum_{m,n=0}^{\infty}A^{\pm}_{mn}|m\rangle_{a1}|n\rangle_{a2}|n\rangle_{b1}|m\rangle_{b2},\nonumber\\
|\Phi^{(\pm)}_{mac}\rangle=\sum_{m,n=0}^{\infty}B^{\pm}_{mn}|m\rangle_{a1}|m\rangle_{a2}|n\rangle_{b1}|n\rangle_{b2}, \label{state_FE}
\end{eqnarray}
where $A^{\pm}_{mn}$ and $B^{\pm}_{mn}$ are coefficients depending on the parametric gain~\cite{Simon}. One can see that there is exact photon-number correlation between orthogonally polarized modes in the case of $|\Psi^{(\pm)}_{mac}\rangle$ states and between similarly polarized modes in the case of $|\Phi^{(\pm)}_{mac}\rangle$ states.

\begin{figure}[h]
\begin{center}
\includegraphics[width=0.35\textwidth]{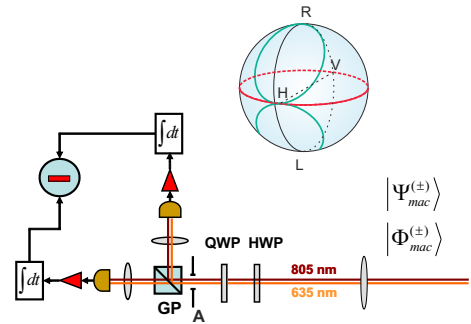}
\caption{Setup for measuring the polarization properties of macroscopic Bell states and the Poincare sphere with the `trajectories' accessible by rotating only the HWP (red line) and only the QWP (green line).}\label{setup}
\end{center}
\end{figure}

Not dwelling on the generation setup, which was described in detail in Ref.~\cite{Macrobell}, we will consider here only the measurement part (Fig.~\ref{setup}). The state under study is filtered in the transverse wavevector (angle of scattering) by placing  a lens (focal length $30$ cm) and a circular aperture (A) with the diameter $9$ mm in its focal plane. This angular filtering is lossless and also provides restriction of the frequency spectrum~\cite{Macrobell}, which then consists of two peaks, centered at wavelengths $635$ and $805$ nm and having widths about $20$ nm each. The whole beam, containing both wavelengths, is sent to the Stokes-measurement setup, including a Glan prism (GP) preceded by a halfwave plate (HWP) and a quarter-wave plate (QWP). The plates are zero-order ones, hence they provide nearly the same phase shifts for both wavelengths. The HWP oriented at an angle $\chi_H$ and a QWP oriented at an angle $\chi_Q$ select a direction in the Stokes space with the coordinates
\begin{eqnarray}
\theta=\arccos[\cos(2\chi_Q)\cos(4\chi_H-2\chi_Q)],\nonumber\\
\phi=-\arctan[\tan(2\chi_Q)/\sin(4\chi_H-2\chi_Q)], \label{angles}
\end{eqnarray}
which define the $\mathbf{n}$ vector.
The Stokes observable $S_{\mathbf{n}}$ is then given by the difference of the intensities in the two output channels of the Glan prism. Note that the beam of squeezed vacuum is analyzed as a whole, without separating the wavelengths or selecting single longitudinal or transverse modes. The intensities are measured pulse by pulse, using PIN diodes followed by charge-integrating amplifiers~\cite{Iskhakov}. The mean value of the intensity difference (found by averaging over $20000$ pulses) defines then $\langle S_{\mathbf{n}}\rangle$ (in arbitrary units), and the variance corresponds to $\Delta S^2_{\mathbf{n}}$. Further analysis of the Stokes observable distribution allows one to perform polarization tomography of the radiation under study~\cite{Karmasexp,poltomo}, but this is outside of the scope of this paper.

\begin{figure}[h]
\begin{center}
\includegraphics[width=0.5\textwidth]{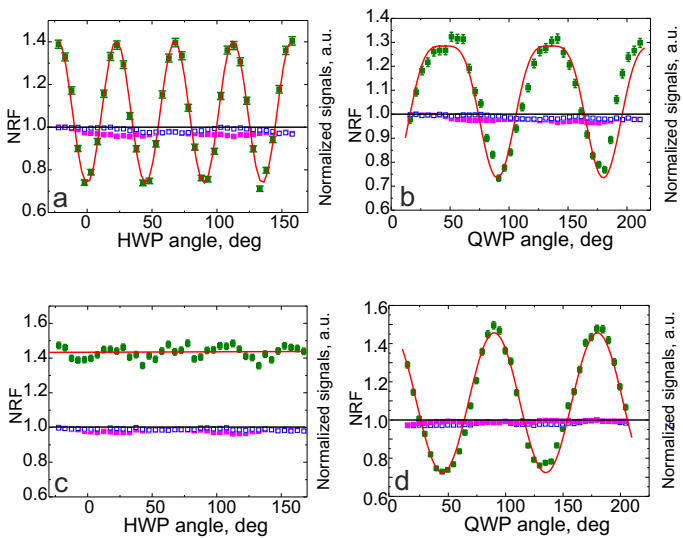}
\caption{Normalized variances of the Stokes observable $\hat{S}_{\mathbf{n}}$ with $\mathbf{n}$ determined by the rotation of a HWP (\textbf{a,c}) and a QWP (\textbf{b,d}) for the $|\Psi^{(+)}_{mac}\rangle$ (\textbf{a,b}) and
$|\Phi^{(+)}_{mac}\rangle$ (\textbf{c,d}) states. Red solid lines show the theoretical fit according to Eqs.~(\ref{PsiHWP},\ref{PhiHWP}). Blue and red points show the normalized signals in both channels.}\label{all}
\end{center}
\end{figure}

In the first experimental run, we only rotated the HWP and QWP separately. The corresponding trajectories made on the Poincare sphere are shown in Fig.~\ref{setup}. The mean values of the intensities in both channels remained constant and equal~\cite{Macrobell} while the variance measurement revealed modulation. As an example, Fig.~\ref{all} shows the normalized variance of the Stokes observable for the $|\Psi_{mac}^{(+)}\rangle$ and $|\Phi_{mac}^{(+)}\rangle$ states versus the orientations of the HWP and QWP. The variance is normalized to the mean value $\langle S_0\rangle$, which corresponds to the shot-noise level~\cite{Iskhakov}. The resulting quantity is called noise reduction factor (NRF), $NRF\equiv\Delta S^2_{\mathbf{n}}/\langle S_0\rangle$, and shows the degree of two-mode squeezing.

Solid lines in Fig.~\ref{all} show the fit according to the formulas
\begin{eqnarray}
NRF(\chi_H)=1+\eta N-\eta (1+N)\cos(8\chi_H),\nonumber\\
NRF(\chi_Q)=1+\eta N+\frac{\eta(N+1)}{4}\nonumber\\
\times(1-4\cos(4\chi_Q)-\cos(8\chi_Q))
\label{PsiHWP}
\end{eqnarray}
for $|\Psi_{mac}^{(+)}\rangle$, and
\begin{eqnarray}
NRF(\chi_H)=1+\eta+2\eta N,\nonumber\\
NRF(\chi_Q)=1+\eta N+\eta (1+N)\cos(4\chi_Q)
\label{PhiHWP}
\end{eqnarray}
for $|\Phi_{mac}^{(+)}\rangle$. Equations (\ref{PsiHWP},\ref{PhiHWP}) were calculated theoretically from the variances of the Stokes observables $S_{1,2,3}$~\cite{Macrobell} and transformations (\ref{angles}). The finite detection efficiency of the setup $\eta$ was taken into account using the beamsplitter model. It is worth mentioning that, strictly speaking, this model describes only part of the detection flaws, namely, the non-unity quantum efficiency of the detectors and the optical losses. Another important source of noise is the mode mismatch~\cite{two-color}, due to the fact that the radiation at both wavelengths, $805$ nm and $635$nm, is selected by the same aperture. The mode mismatch adds some value to NRF and, in principle, can make NRF higher than unity even in the case of ideal two-mode squeezing. In our experiment, due to the small parametric gain, the NRF increase was small and could be described by a reduced value of quantum efficiency. The parameter $N$ is the mean photon number per mode, equal to $N=\sinh^2\Gamma$.  Note that the fit, with the fitting parameters $N,\eta$, is rather sensitive to the value of $\eta$, which is thus estimated as $\eta=0.26\pm0.02$, but gives just a rough estimate for the photon number: $N=0.2\pm0.1$.

The results show strong dependence of $\Delta S^2_{\mathbf{n}}$ on the orientation of $\mathbf{n}$. The exception is the case (c), which shows $\Delta S^2_{\mathbf{n}}$ as a function of the HWP orientation for $|\Phi^{(+)}_{mac}\rangle$. This is in agreement with the fact that, similarly to the two-photon Bell state $|\Phi^{(+)}\rangle$, its macroscopic analogue $|\Phi_{mac}^{(+)}\rangle$ is invariant to linear polarization rotation. Rotation of the QWP, though, reveals some modulation of the measured variance (Fig.\ref{all}d). The fact that the variance of the Stokes observable is sensitive to polarization transformations while its mean value is invariant to it, indicates the presence of hidden polarization. Similar behaviour is observed for the $|\Phi_{mac}^{(-)}\rangle$ state.

\begin{figure}[h]
\begin{center}
\includegraphics[width=0.23\textwidth]{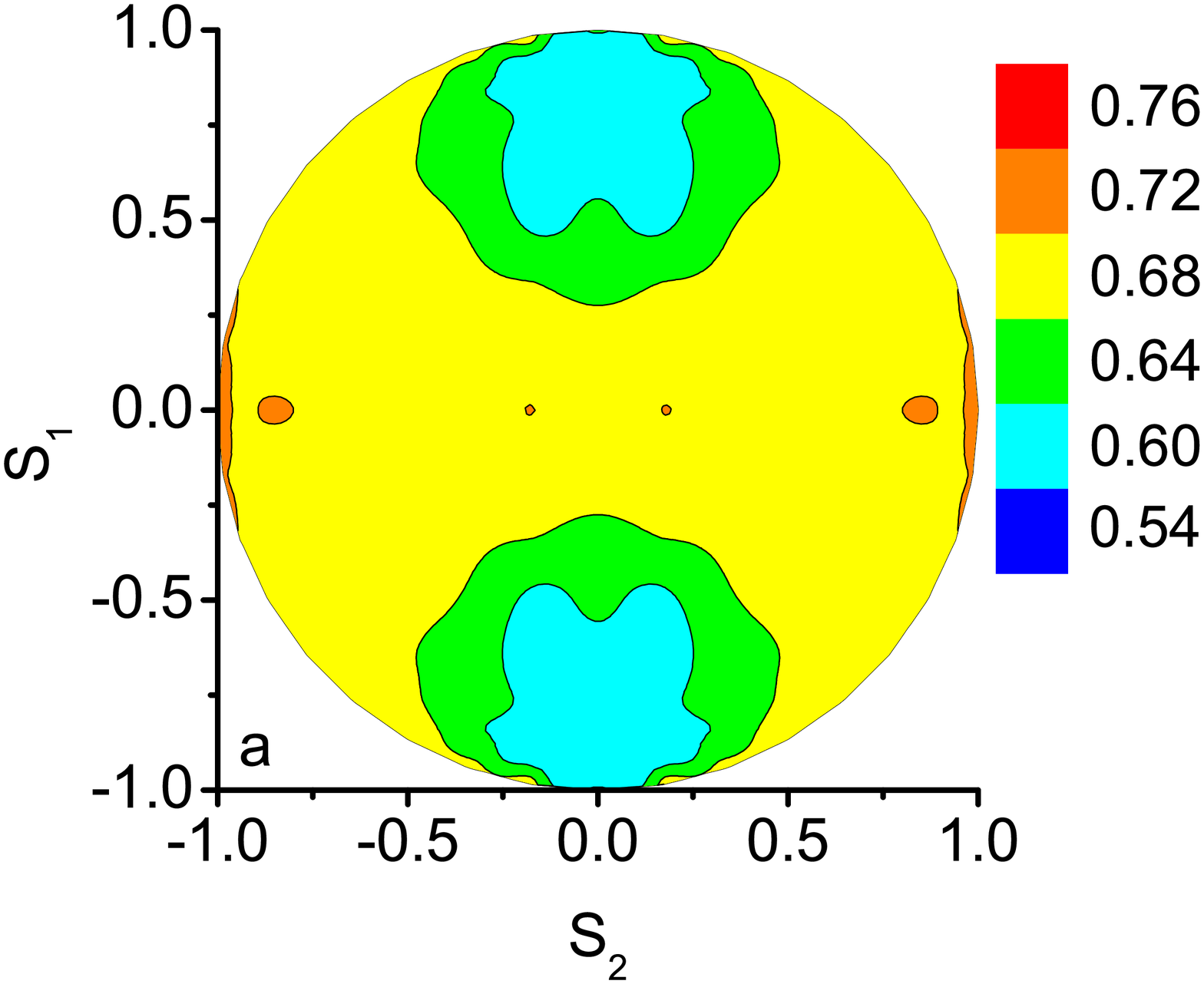}
\includegraphics[width=0.23\textwidth]{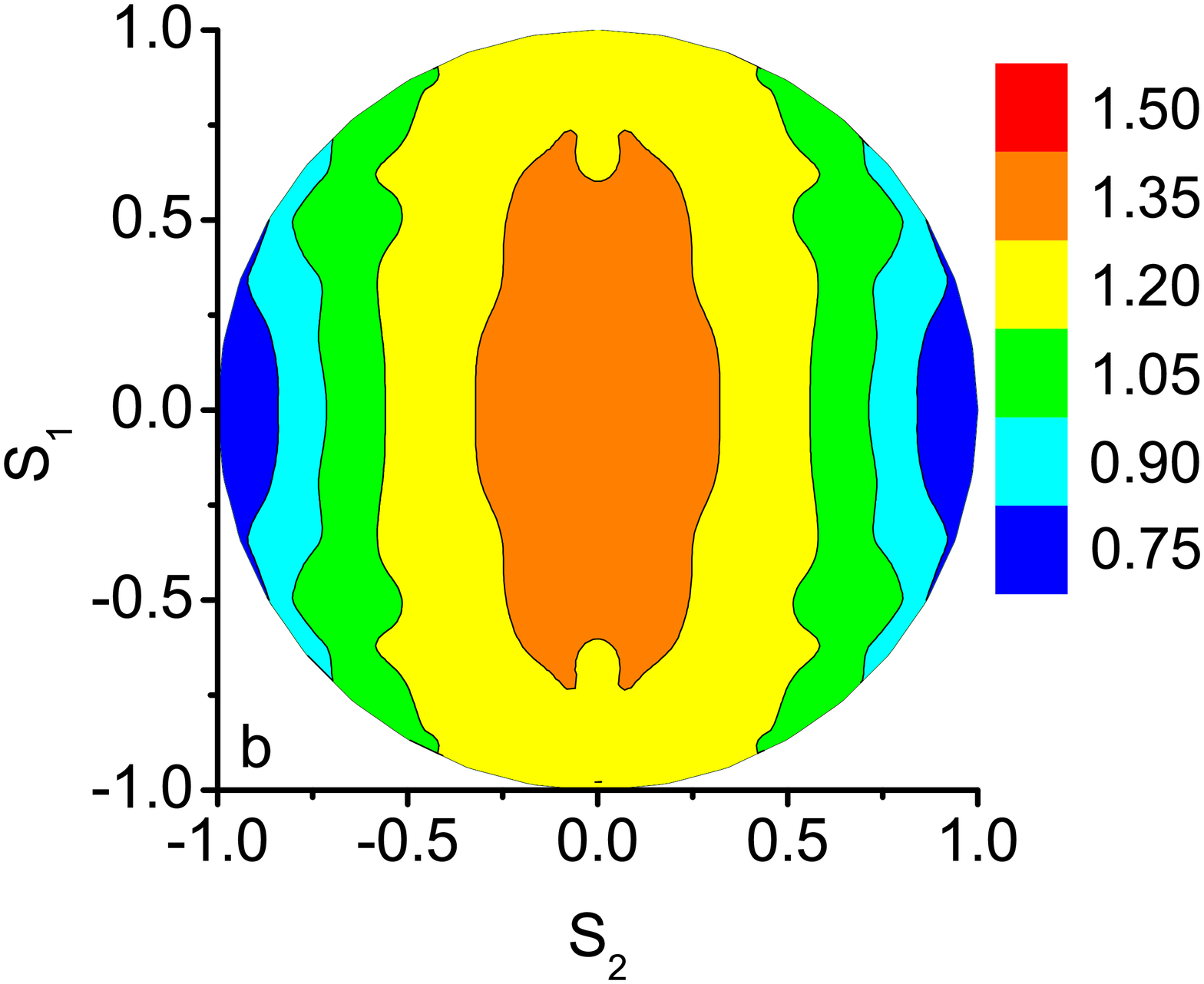}
\caption{Normalized variance $\hbox{Var}S_{\mathbf{n}}/\langle S_0\rangle$ for the macroscopic singlet state $|\Psi^{(-)}_{mac}\rangle$ (a) and one of the triplet states, $|\Psi^{(+)}_{mac}\rangle$ (b), depicted by color on a sphere showing $\mathbf{n}$ orientation. For simplicity, the sphere is projected on the ($S_2$, $S_1$ plane). Note that the color code range is chosen differently for a and b in order to properly show the data.}\label{spheres}
\end{center}
\end{figure}

At the same time, no hidden polarization was observed for the macroscopic singlet state $|\Psi_{mac}^{(-)}\rangle$~\cite{Macrobell}. Both mean value and the variance of the Stokes observable were invariant to the HWP and QWP positions, changed separately. To complete these data by choosing all possible HWP and QWP orientations, we performed an experiment where the plates were moved in steps ($2.5^{\circ}$ for HWP and $5^{\circ}$ for QWP) spanning approximately one octant of the Poincare sphere. The results are shown in Fig.~\ref{spheres}: the normalized variance of the Stokes observable, for the states $|\Psi^{(-)}_{mac}\rangle$ (a) and $|\Psi^{(+)}_{mac}\rangle$ (b). The value of NRF is depicted in color on a unit sphere, which denotes the $\mathbf{n}$ orientation. The figure shows  the projection of the sphere on the $(S_2,S_1)$ plane.
One can see that the distribution has much more pronounced modulation for the triplet state than for the singlet one. Theoretically, the singlet state should show no modulation at all; the small modulation seen in Fig.~\ref{spheres}a is apparently caused by the non-ideal preparation of the state.

From these data, one can calculate the values of the first-order DP, according to definition (\ref{DP1}), and for the second-order DP, according to definition~(\ref{DP_Klyshko}). The results are shown in the second and third columns of Table~\ref{table1}.

\begin{table}[h]
\caption{First-order, second-order, and fourth-order DP calculated for the singlet macroscopic Bell state and for one of the triplet macroscopic Bell states.}\label{table1}
\begin{center}
\begin{tabular}{|p{0.15\linewidth}|p{0.27\linewidth}p{0.27\linewidth}p{0.27\linewidth}|}

\hline State  & $P_1$ & $P_2$ & $P_4$ \\
\hline
$|\Psi^{(-)}_{mac}\rangle$          & $0.016\pm0.001$    & $0.14\pm0.01$               & $0.28\pm0.05$         \\
$|\Psi^{(+)}_{mac}\rangle$          & $0.019\pm0.001$    & $0.36\pm0.01$               & $0.64\pm0.02$   \\
\hline
\end{tabular}
\end{center}
\end{table}

The second-order polarization degree (\ref{DP_Klyshko}) to be observed for the triplet states, from Eqs.~(\ref{PhiHWP}) or Eqs.~(\ref{PsiHWP}), is
\begin{equation}
P_2^{\mathrm{trip}}=\frac{\eta(1+N)}{1+\eta N}.
\label{DP_2_th}
\end{equation}
It differs from unity only due to the non-ideal quantum efficiency $\eta<1$, which leads to the mixing of the state with the vacuum.  Note that the DP gets close to unity at high parametric gain, $N\gg1$, regardless of the quantum efficiency. For the singlet state, the second-order DP, theoretically, should be zero at any parametric gain and at any quantum efficiency:
\begin{equation}
P_2^{\mathrm{sing}}=0.
\label{DP_2_sin}
\end{equation}

Alternatively, one can define the second-order DP in terms of normally ordered correlation functions~\cite{Klyshko}; this will provide the invariance of the DP to losses or non-ideal quantum efficiency. On the other hand, in the relatively high-gain regime used in our experiment, normally ordered correlation functions are almost insensitive to polarization transformations. The reason is that they contain high background, which is determined by only the intensities measured by the detectors and hence independent on the polarization transformations.

The operational definition (\ref{DP_Klyshko}) of the DP as the visibility of the second-order polarization pattern admits a simple extension to the arbitrary-order case:
\begin{equation}
P_k\equiv\frac{\hbox{supp}_{\mathbf{n}}\Delta S^k_{\mathbf{n}}-\hbox{inf}_{\mathbf{n}}\Delta S^k_{\mathbf{n}}}{\hbox{supp}_{\mathbf{n}}\Delta S^k_{\mathbf{n}}+\hbox{inf}_{\mathbf{n}}\Delta S^k_{\mathbf{n}}},
\label{DP_k}
\end{equation}
where $\Delta S^k_{\mathbf{n}}$ is the $k$th-order central moment, $\Delta S^k_{\mathbf{n}}\equiv\langle(S_{\mathbf{n}}-\langle S_{\mathbf{n}}\rangle)^k\rangle$.

\begin{figure}[h]
\begin{center}
\includegraphics[width=0.23\textwidth]{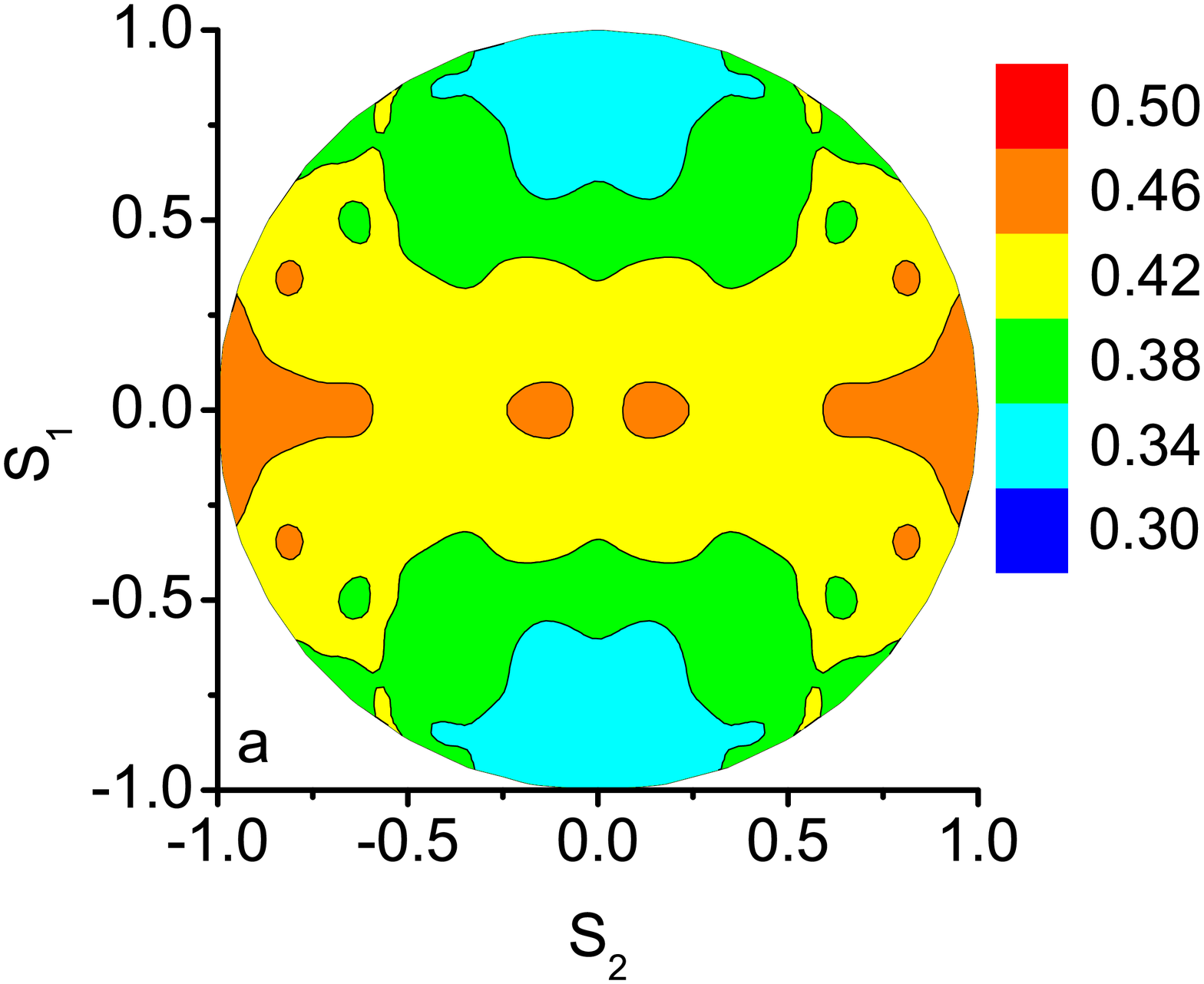}
\includegraphics[width=0.23\textwidth]{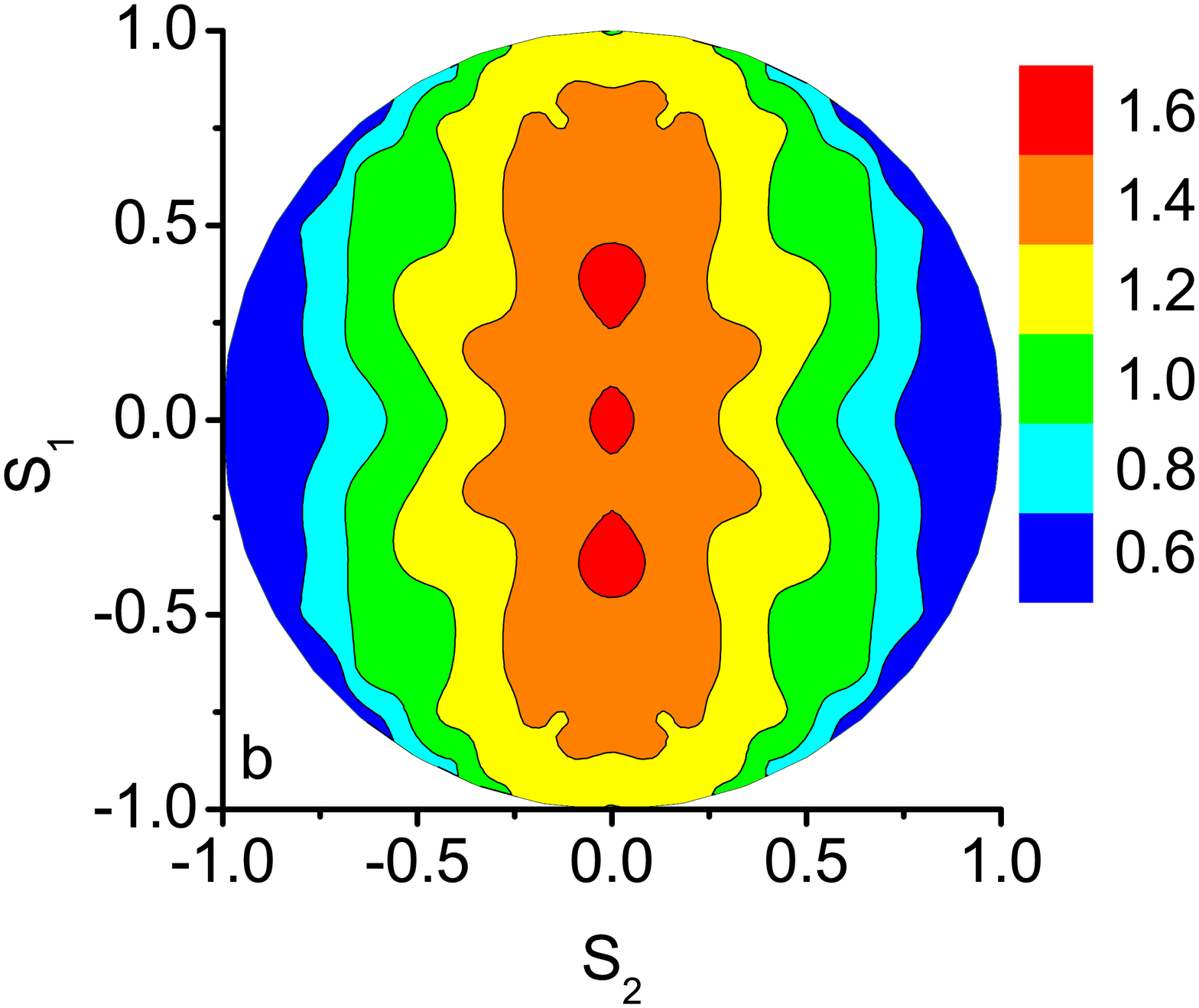}
\caption{Fourth-order central moment $\Delta S_{\mathbf{n}}^4$, normalized to its value for a coherent state with the same intensity, for the macroscopic singlet state $|\Psi^{(-)}_{mac}\rangle$ (a) and one of the triplet states, $|\Psi^{(+)}_{mac}\rangle$ (b).}\label{fourth}
\end{center}
\end{figure}

For both singlet and triplet macroscopic Bell states, the third central moments of all Stokes observables are equal to zero, as all Stokes histograms are symmetric. Moreover, these histograms are Gaussian, according to the theory, hence all even-order central moments can be expressed in terms of the variance. However, to demonstrate the definition of the higher-order DP (\ref{DP_k}), we have calculated the fourth-order moments from the histograms of the Stokes observables. The contribution of the electronic noise was measured independently, from the distribution obtained without any incident light, and eliminated numerically.
The resulting distributions of the fourth-order moments are shown in Fig.~\ref{fourth}, again, by color coding, and
the orientation of $\mathbf{n}$ is shown on a sphere projected onto the $(S_2,S_1)$ plane.
The obtained values of the fourth-order DP are given in Table~\ref{table1} (fourth column).
The fourth moment was normalized to its value for a coherent state with the same intensity.
As expected, the singlet state manifests much smaller modulation of the fourth-order moment
than the triplet state. Also, the fourth moment is
suppressed for the singlet state compared to its value for a coherent state.

In conclusion, we have demonstrated the `hidden polarization' effect for the triplet macroscopic Bell states and found the second-order degree of polarization for such states. Moreover, we have extended the Klyshko definition of second-order degree of polarization to the higher-order case; the fourth-order degree of polarization was calculated for the macroscopic Bell singlet and triplet states. Certainly, for the class of Gaussian states, which includes macroscopic Bell states studied in our experiment, all higher-order moments are determined by the first two ones. However, the concept of higher-order degree of polarization can be very useful in the general case of non-Gaussian states.

We acknowledge the financial support of the European Union under project COMPAS No.
212008 (FP7-ICT) and of the Russian Foundation for Basic Research, grant 10-02-00202.
T.Sh.I. acknowledges funding from Alexander von Humboldt Foundation.


\begin{references}
\bibitem{second-order} There is ambiguity in using the terms 'second-order' and 'fourth-order' in quantum optics. Here, by 'second-order', we will mean 'second-order in the intensity', i.e., fourth-order in the field.
\end{references}
\end{document}